\documentclass[epsfig,12pt,onecolumn]{article}
\usepackage{amsfonts}
\usepackage{amssymb}
\usepackage{amsmath}
\usepackage{multicol}
\usepackage{graphicx}
\usepackage{float}
\usepackage{caption}

\input{tcilatex}
\makeatletter \@addtoreset{equation}{section}

\def \be{\begin{equation}}
\def \ee{\end{equation}}
\def \bea{\begin{eqnarray}}
\def \eea{\end{eqnarray}}

\newcommand{\nc}{\newcommand}
\nc{\al}{\alpha} \nc{\bib}{\bibitem} \nc{\la}{\lambda}
\nc{\C}{\mbox{\hspace{1.24mm}\rule{0.2mm}{2.5mm}\hspace{-2.7mm} C}}
\nc{\R}{\mbox{\hspace{.04mm}\rule{0.2mm}{2.8mm}\hspace{-1.5mm} R}}

\begin{document}

\title{Unitary description of the black hole by prime numbers}
\author{M. Bousder$^{1}$\thanks{%
mostafa.bousder@um5.ac.ma } \\
$^{1}${\small LPHE-MS Laboratory, Department of physics,}\\
\ {\small Faculty of Science, Mohammed V University in Rabat, Rabat, Morocco}%
}
\maketitle

\begin{abstract}
In this paper, we study the thermofield double states of doubly-holographic
gravity in two copies of the horizons. We show that the asymptotically AdS
spacetimes describe an entangled states of a pair of CFTs based on the Farey
sequence. We propose a new technique to geometrize the black hole horizon.
Our protocol is based on the so-called Farey diagram. We construct states
and entropies to describe the unit cells on the horizon. As a result, we
have proved that the quantum states on the horizon are encoded by prime
numbers. Therefore, we found that the entropy of the code space and area law
are writtens in logarithmic form of the prime numbers. We show that the
number of connected components of the Farey sequence can build the
Fermi--Dirac distribution. To solve the information paradox problem, we find
that the Hawking radiation follows geodesic of the Farey diagram, then he
turns around and falls on the horizon. Our aim is to show that there is
appearance of several Page times, because of discontinuous emission of these
radiations. Finally, we mention the possibility to describe the quantum Hall
effect by using the Farey diagram. In this paper, we find a link between
quantum information and the theory of numbers passing through geometry.

\textbf{Keywords:} Thermofield double, Black hole, Information paradox,
Entanglement entropy.
\end{abstract}

\section{\textbf{Introduction}}

(i) \textit{Background. }\newline
In recent breakthrough works, the loss of information during the evaporation
of black holes is an open problem \cite{K}. Many proposals were suggested to
resolve the information paradox. In a lines of the introduction we discuss
the attempts to solve this problem:\newline
$\bullet $ In $AdS/CFT$ \cite{K0}, the $d+1$-dimensional quantum gravity
(AdS) is dual to $d$-dimensional conformal field theory (CFT) which lives on
the asymptotic boundary of space-time. The $AdS/CFT$ correspondence shows
that information can escape from a black hole. The unitarity of the CFT
implies the information reservation, so, the black hole evaporation process
is expected to be unitary \cite{K1}. In this case, the evaporation of the
black hole has a dual unitary description in the CFT.\newline
$\bullet $ The so-called Page curve as a function of time describes the
unitary process of black hole evaporation \cite{K2,K3}. In this curve, the
entropy increases until the Page time, then it begins to decrease. In
Jackiw-Teitelboim (JT) gravity in $AdS_{2}$, the Page curve was reproduced
explicitly without assuming the unitarity. At the Page time, the
entanglement entropy in the black hole transition marks a change between
initial and final states.\textbf{\ }Before the transition, there is a simple
spatial division between the radiation and the degrees of freedom of the
black holes. After this transition, the interior region of the black hole
covered by the island structure \cite{K5}. In \cite{K6}, the entanglement
wedge reconstruction occurs in the island, i.e. inside the black hole
"interior".\newline
$\bullet $ In \cite{ER3}, the Page curve is reproduced according to the
Ryu-Takayanagi (RT) formula for the von Neumann entropy of radiation, which
was inspired by the quantum extremal surface formula (QES) for the
holographic entanglement entropy. The entanglement entropy (EE) of a
boundary region is given by the area of RT surfaces, corresponding to the
bulk minimal surface. The thread prescription which is equivalent to the
quantum extremal surface (QES) prescription \cite{Q2,Q3}.\newline
$\bullet $ The ER = EPR \cite{ER0} has given a new perspective on the
gauge/gravity correspondence. Under this paradigm, a pair of entangled black
holes are joined by an Einstein-Rosen (ER) bridge. This suggests that there
is a relation between quantum entanglement and geometric spatial. \newline
$\bullet $ The construction of multi-boundary or wormholes was applied in
the framework of the End-of-the-World (EOW) brane model \cite{K4} in three
dimensions and without quantum fields.\newline
$\bullet $ Recent work has focused on the issue of the thermofield double
(TFD) states is a central ingredient. The connection between boundary
information-theoretic quantities and bulk areas has been extended to a
conjectured \cite{QQ1}. In parallel, there have also been numerous proposals
to directly modeling the black hole (BH) evaporation and calculations of the
entanglement entropy \cite{ER5}.\newline
$\bullet $ In \cite{SQ}, the black hole AdS geometry is coupled to a bath
which absorbs the black hole radiation, also the Page curve is characterized
by the unitary black hole evaporation.

(ii) \textit{Goal and strategy.}

In this work, we presente a preparation scheme for Farey sequence that can
be implemented on a $CFT_{d}$ on the boundary asymptotically $AdS_{d+1}$
boundary of the dual gravity theory. Our first goal then is to figure out
what the relation is between the TFD state the geometry of the region near
hrizon and. Our second goal is to determine the black hole statistic by
using the TFD states. In this basis, we imagine a temperature at the horizon
tends to $\infty $. Then write the states $\left\vert TFD\right\rangle $ as
a function of the states $\left\vert TFD(\beta =0)\right\rangle $. This new
technique will allow us to find another path to determine both the geometry,
the state encoding and the statistics of the black hole. By studying the the
number of BH states, we can understand that the BH undergoes the
Fermi--Dirac particle-energy statistics.\newline
To study the entanglement entropy between the BH radiation and the quantum
state associated to the remaining BH in the framework of the Page curve. We
propose a new description which begins with the study of TFD states in the
horizon. Then we propose a particular operator which can destroy the
entanglement at the intinit temperature, and we write TFD states in terms of
this operator. We assume that the destruction of the entanglement is done in
the time of Page. Our aim is to geometrize the AdS background of the BH in
the presence of entangled states. When we remove the entanglement by the
operator, then we build the entanglement by the results of calculations, we
find the AdS geometry dual to CFT of the entangled states. This geometry
allows to see the entanglement as a geodesic in the background. Our results
show that there is not a loss of Hawking radiations in space-time, but the
radiations follow a special geodesic generated by the Farey diagram lines.

(iii)\textit{\ Organization.}\newline
The paper is organised as follows. In Section 2 we introduce and review the
TFD with an operator that removes entanglement for an infinite temperature ($%
T\rightarrow \infty $). Then in Section 3 wee discuss the AdS background
geometry by the description of the CFT states at $T\rightarrow \infty $,
from the point of view of the Farey diagram. Following that, in Section 4 we
build microstates of unitary describtion in Farey diagram (AdS geometry),
and we find a connection between multi-boundary and microstates. In Section
5 we find that the Fermi-Dirac statistics governs unitary processes of the
system. Using the black hole first law of thermodynamics in Section 6 we
find a quantum description of the CFT states by the quantum hall effect. In
Section 7 we present some conclusions and outlook.

\section{Left and right horizons in CFT}

(i) \textit{Motivation}\newline
Let $\emph{H}=\emph{H}_{L}\otimes \emph{H}_{R}$ be the Hilbert space of the
full $CFT$ and $\left\vert u\right\rangle $ denote the quantum state of a
system with $N$ particles. Hamiltonian evolution is generated by \emph{H}$%
_{TFD}=\emph{1}_{L}\otimes \emph{H}_{R}+\emph{H}_{L}\otimes \emph{1}_{R}$.
Let us define the thermofield double (TFD) state as a particular entangled
state \cite{ER1,ER4}:
\begin{equation}
\left\vert TFD\right\rangle =\frac{1}{\sqrt{Z\left( \beta \right) }}%
\sum_{j}e^{-\beta E_{j}/2}\left\vert u_{j}\right\rangle _{L}\otimes
\left\vert \bar{u}_{j}\right\rangle _{R},  \label{TFD}
\end{equation}%
where $Z$ is the partition function of the CFT at temperature $T=\beta ^{-1}$
and $e^{-\beta E_{j}/2}/\sqrt{Z\left( \beta \right) }$ is a normalization
factor such that $_{k}\left\langle TFD\right. \left\vert TFD\right\rangle
_{k}=1$. The TFD state was suggested outside of the black holes. In this
last equation, we consider $t$ as a parameter labeling time-dependent TFD
state at a common instant $\left\vert TFD\left( t\right) \right\rangle
=e^{it\left( H_{L}+H_{R}\right) }\left\vert TFD\right\rangle $. In each of
these states at time $t$, there exists a projection operator $%
P_{t}=\left\vert TFD(t)\right\rangle \left\langle TFD(t)\right\vert $.
According to \cite{ER0}, two entangled states with different values of $t$
are linked by the forward time evolution on the two sides of \ $CFT$. Note
that the projection operator is expressed in terms of $t$. $\left\vert
u_{j}\right\rangle _{LR}$ defined in the microscopic UV-complete theories,
they create $CFT$ states with energy \cite{ER01}. By tracing out the
exterior states in one horizon we obtain the density matrix. The mixed state
given by the incoherent sum over all generalized TFD states $\rho
_{TMD}=\left( 1/N\right) \sum_{j}\left\vert TFD\right\rangle _{j\text{ \ }%
j}\left\langle TFD\right\vert $ call the the thermo-mixed double, where $N$
the total number of basis states. The thermal density matrix as arising from
entanglement. The corresponding bulk geometry of the wormhole is formed by $%
tr\left( \rho _{TMD}^{2}\right) =\sum_{j}e^{-2\beta E_{j}}/Z^{2}$, where $%
Z=\sum_{j}e^{-\beta E_{j}}$ is partition function, with the square number $2$
corresponds to the two horizons.

We consider the TFD of dual to the left and right sides of the geometry by
two copies of the horizons $\mathcal{H}_{L}$(left horizon) and $\mathcal{H}%
_{R}$(right horizon) in CFT. The eigenstate $\left\vert u_{j}\right\rangle
_{L}$ (or $\left\vert \bar{u}_{j}\right\rangle _{R}$) of the $CFT_{L}$ (or $%
CFT_{R}$) correspondes to the degrees of freedom in $\mathcal{H}_{L}$ (or $%
\mathcal{H}_{R}$). In the context of holographic gravity, the two sides of \
$CFT$ are connected by a wormhole, or ER bridge, according to the $TFD$
states Eq.(\ref{TFD}). The $TFD$ is a maximally entangled state, which
represents the formal purification of the thermal mixed state of a one
horizon ($\mathcal{H}_{L}$ or $\mathcal{H}_{R}$) \cite{ER2}, with the
reduced density matrix within $AdS/CFT$ via the Ryu-Takayanagi formula \cite%
{ER3}.

(ii) \textit{Strategy}\newline
To study TFD states at the time of entanglement removal at an infinite
temperature $\left( \beta =0\right) $ in the horizon. We start by define an
entangled state which is expressed as
\begin{equation}
\left\vert \psi \right\rangle =\left\vert TFD\left( \beta =0\right)
\right\rangle _{LR}+\left\vert TFD\left( \beta =0\right) \right\rangle
_{RL}\in H.  \label{F0}
\end{equation}%
At this point, the entangled state can be decomposed into a $LR$ and $RL$
part as follows
\begin{equation}
\left\vert \psi \right\rangle =\frac{1}{\sqrt{Z\left( \beta \right) }}\left(
\sum_{j=1}^{N_{-}}\left\vert u_{j}\right\rangle _{L}\otimes \left\vert \bar{u%
}_{j}\right\rangle _{R}+\sum_{k=1}^{N_{+}}\left\vert \bar{u}%
_{k}\right\rangle _{R}\otimes \left\vert u_{k}\right\rangle _{L}\right) ,
\label{s1}
\end{equation}%
Here, $N$ denotes the number of the quantum state, where $N=N_{-}+N_{+}$,
with the number $N_{-}$ and $N_{+}$ of copies of the states $\left\vert
u\right\rangle $ and $\left\vert \bar{u}\right\rangle $, respectively, and $%
\left\vert u_{j}\right\rangle \neq \left\vert u_{j+1}\right\rangle $. We
define an unitary operator $\mathcal{W}$ which verifies
\begin{eqnarray}
\mathcal{W}\left\vert u_{j}\right\rangle _{L} &=&\left( -1\right)
^{j-1}\left\vert u\right\rangle _{L},  \label{s3} \\
\mathcal{W}\left\vert \bar{u}_{j}\right\rangle _{R} &=&\left( -1\right)
^{j-1}\left\vert \bar{u}\right\rangle _{R},  \notag \\
\mathcal{W}^{2} &=&1,  \notag
\end{eqnarray}%
the aim of this operator is to remove the interaction between states near
horizon at $\beta =0$. We find
\begin{equation}
\mathcal{W}\left\vert \psi \right\rangle \sim \sum_{j=1}^{N_{-}}\left(
-1\right) ^{j-1}\left\vert u\right\rangle \left\vert \bar{u}\right\rangle
+\sum_{k=1}^{N_{+}}\left( -1\right) ^{k-1}\left\vert \bar{u}\right\rangle
\left\vert u\right\rangle ,  \label{s4}
\end{equation}%
which leads to\newline
\textit{If } $N_{-}$\textit{\ is even and }$N_{+}$\textit{\ is even}, one
obtain $\mathcal{W}\left\vert \psi \right\rangle =0$. \newline
\textit{If }$N_{-}$\textit{\ is even and }$N_{+}$ \textit{is odd}, one get
the pure state\ $\mathcal{W}\left\vert \psi \right\rangle \sim \left\vert
\bar{u}\right\rangle \left\vert u\right\rangle $.\newline
\textit{If }$N_{-}$\textit{\ is odd\ and }$N_{+}$ \textit{is even}, one get
the pure state\ $\mathcal{W}\left\vert \psi \right\rangle \sim \left\vert
u\right\rangle \left\vert \bar{u}\right\rangle $.\newline
\textit{If }$N_{-}$\textit{\ is odd\ and }$N_{+}$ \textit{is odd}, one get
the entangled state\ $\mathcal{W}\left\vert \psi \right\rangle \sim
\left\vert u\right\rangle \left\vert \bar{u}\right\rangle +\left\vert \bar{u}%
\right\rangle \left\vert u\right\rangle $.\newline
\textit{If }$N\ $\textit{is infinite}, one get the entangled state\ $%
\mathcal{W}\left\vert \psi \right\rangle _{\infty }\sim \frac{1}{2}\left(
\left\vert u\right\rangle \left\vert \bar{u}\right\rangle +\left\vert \bar{u}%
\right\rangle \left\vert u\right\rangle \right) $, ($\zeta (0)=1/2$, the
analytic continuation of the Riemann zeta function).

\textit{(iii) Results}

We can map initial state $\left\vert \bar{u}\right\rangle \left\vert
u\right\rangle $ and final state $\left\vert u\right\rangle \left\vert \bar{u%
}\right\rangle $ in the CFT, then we can compute the entanglement between
the two states in the CFT. This procedure can represent by the following
quantification. The states $\left\vert 0\right\rangle $ and $\left\vert \bar{%
0}\right\rangle $ are the Rindler vacuum states in the two exterior region
horizons in the right and left horizons, respectively. The state $\left\vert
0\right\rangle \left\vert \bar{0}\right\rangle $ is the outside horizons
vacuum states. \newline
The state $\mathcal{W}\left\vert \psi \right\rangle $ is transformed to%
\begin{equation}
\mathcal{W}\left\vert \psi \right\rangle \sim \alpha \left\vert
u\right\rangle \left\vert \bar{u}\right\rangle +\beta \left\vert \bar{u}%
\right\rangle \left\vert u\right\rangle ,  \label{s5}
\end{equation}%
with%
\begin{equation}
\left( \alpha ,\beta \right) =\left\{
\begin{array}{c}
\left( 0,0\right) \\
\left( 0,1\right) \\
\left( 1,0\right) \\
\left( 1/2,1/2\right) \\
\left( 1,1\right)%
\end{array}%
\right. ,  \label{QQ}
\end{equation}%
such that $\left( \alpha ,\beta \right) \geq \left( 0,0\right) $ are quantum
numbers. The results of this section suggest that $\left\vert
TFD\right\rangle =\mathcal{W}\left( \sum_{j}\sqrt{p_{j}}e^{i\omega
_{j}}\left\vert \psi \right\rangle _{j}\right) $. We can reexpress the TFD
states in terms of the ($\beta =0$) modes as
\begin{equation}
\left\vert TFD\right\rangle =\frac{1}{\sqrt{Z\left( \beta \right) }}%
\sum_{j=1}^{N}e^{-\beta E_{j}/2}\mathcal{W}\left( \left\vert TFD_{j}\left(
\beta =0\right) \right\rangle _{LR}+\left\vert TFD_{j}\left( \beta =0\right)
\right\rangle _{RL}\right) .  \label{F1}
\end{equation}%
As previously mentioned, we adopt the states $\left\vert TFD\left( \beta
=0\right) \right\rangle $ to evaluate the geometric approach of two copies
of the horizons in the holographic CFT. The volume of this geometry
increases in time according to the number of degrees of freedom of the
boundary theory. Note that we can rewrite the state $\left\vert \psi
\right\rangle $ as
\begin{equation}
\left\vert \psi \right\rangle \sim N_{+}\left\vert u\right\rangle \left\vert
\bar{u}\right\rangle +N_{-}\left\vert \bar{u}\right\rangle \left\vert
u\right\rangle .  \label{s6}
\end{equation}%
For entangled states, the left and right systems are identical ($%
N_{-}=N_{+}=N/2$), the state $\left\vert \psi \right\rangle $ can be
represented as $\left\vert \psi \right\rangle \sim \frac{N}{2}\left(
\left\vert u\right\rangle \left\vert \bar{u}\right\rangle +\left\vert \bar{u}%
\right\rangle \left\vert u\right\rangle \right) $. Comparing Eq.(\ref{s1})
to Eq.(\ref{s5}), we conclude that

\begin{equation}
\left\vert u\bar{u}\right\rangle =\frac{N_{-}-\beta }{\alpha -N_{+}}%
\left\vert \bar{u}u\right\rangle .
\end{equation}%
where $\left\vert u\bar{u}\right\rangle =\left\vert u\right\rangle
\left\vert \bar{u}\right\rangle $. Using the normalization: $\left\langle u%
\bar{u}\right. \left\vert u\bar{u}\right\rangle =1$, we notice that $\alpha
+\beta =N$ or $\alpha -\beta =N_{+}-N_{-}$. Then, by applying this
observation, we obtain
\begin{eqnarray}
\left\vert u\bar{u}\right\rangle &=&+\left\vert \bar{u}u\right\rangle \text{
for }\alpha +\beta =N,  \notag \\
\left\vert u\bar{u}\right\rangle &=&-\left\vert \bar{u}u\right\rangle \text{
for }\alpha -\beta =N_{+}-N_{-}.
\end{eqnarray}%
Since the Eq.(\ref{QQ}) satisfy $\alpha +\beta =\left\{ 0,1,2\right\} $ and $%
\alpha -\beta =\left\{ -1,0,1\right\} $, we can reexpress the generator in
terms of the $N$ modes as%
\begin{eqnarray}
\left\vert u\bar{u}\right\rangle &=&+\left\vert \bar{u}u\right\rangle \text{
\ for \ }N=\left\{ 1,2\right\} ,  \label{SS} \\
\left\vert u\bar{u}\right\rangle &=&-\left\vert \bar{u}u\right\rangle \text{
\ for \ }N\geq 2\text{ .}  \notag
\end{eqnarray}%
These two relationships above, implies that the state $\mathcal{W}\left\vert
\psi \right\rangle $ is not an entangled state
\begin{equation}
\left\vert \psi \right\rangle \overset{\text{entanglement removal}}{%
\longrightarrow }\mathcal{W}\left\vert \psi \right\rangle .
\end{equation}%
The operator $\mathcal{W}$ allows the system to be transferred to another
state. In this case we suppose that before the Page time the black hole
exists in the state $\mathcal{W}\left\vert \psi \right\rangle $, after this
time, the outgoing radiation and the quantum state associated to the
remaining black hole are entangled according to the state $\left\vert \psi
\right\rangle $.
\begin{equation*}
\mathcal{W}\left\vert \psi \right\rangle \overset{\text{after Page}\ \text{%
time}}{\longrightarrow }\left\vert \psi \right\rangle .
\end{equation*}%
To describe the Majorana states we can choose $N=2$ for the first case Eq.(%
\ref{SS}). The change of sign $\left( +\right) $ to $\left( -\right) $ in $%
N=2$, lead to additional degeneracy. \newline
Now, we consider $N_{+}=N_{-}=N/2$. The coherent states can be expressed in
terms of Fock states in the standard way as%
\begin{equation}
\left\vert \psi \right\rangle =\frac{1}{\sqrt{Z\left( \beta \right) }}\left[
\sum_{j=1}^{N_{-}}e^{-\beta E_{j}/2}\left\vert u_{j}\right\rangle
_{L}\otimes e^{+\beta E_{j}/2}\left\vert \bar{u}_{j}\right\rangle
_{R}+\sum_{k=1}^{N_{+}}e^{+\beta E_{k}/2}\left\vert \bar{u}_{k}\right\rangle
_{R}\otimes e^{-\beta E_{k}/2}\left\vert u_{k}\right\rangle _{L}\right] .
\end{equation}%
We use the transformation%
\begin{equation}
\sqrt{2}e^{+\beta E_{j}/2}\left\vert \bar{u}_{j}\right\rangle
_{R}\longrightarrow \left\vert \bar{u}_{j}^{\ast }\right\rangle _{R},\text{
\ \ \ \ \ \ \ \ \ }\left\vert u_{k}\right\rangle _{L}\longrightarrow
\left\vert u_{k}^{\ast }\right\rangle _{L}
\end{equation}%
then, the evolution of the horizon geometry is described by\ the following
entangled state%
\begin{equation}
\left\vert \psi \left( t\right) \right\rangle =\frac{1}{\sqrt{2}}\left(
\left\vert TFD^{\ast }\left( t\right) \right\rangle _{LR}+\left\vert
TFD^{\ast }\left( t\right) \right\rangle _{RL}\right) ,  \label{F}
\end{equation}%
where $\left\vert TFD^{\ast }\left( t\right) \right\rangle _{LR}$ is the
evolution of the TFD state in the horizon from $L$ to $R$ at time $t$. If we
compare Eq.(\ref{F}) with Eq.(\ref{F0}), we notice that in the horizon at $%
\beta =0$, there is still an evolution of the system over time. Indeed, the
temperature is infinite only in certain time, but the equivalence between
Eqs.(\ref{F0},\ref{F}), shows that the termperature is infinite in all time:
$\forall t$, $\left\vert TFD\left( \beta =0\right) \right\rangle _{LR}\sim
\left\vert TFD^{\ast }\left( t\right) \right\rangle _{LR}$. This shows that
there is always a production of the Hawking radiation. Therefore, there is
always an entanglement between these radiations and the remaining hole. This
may be due to several effects, but the most realistic effect is that the
radiation comes out of the black hole, after a certain time it follows
spetial geodesic so that they go back and fall on the BH horizon. The
objective in the next section is to determine this geodesic.

\section{Farey diagram of the AdS background}

\begin{figure}[H]
\centering
\includegraphics[width=9cm]{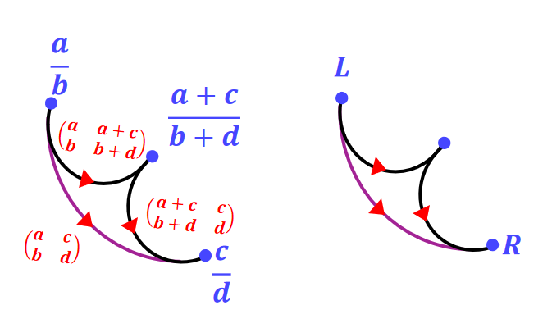}
\caption{To go from$\mathcal{H}_{L}$ to $\mathcal{H}_{R}$, we can use the $%
2\times 2$ matrices correspond to linear transformations, which are formed
by Farey sequence. }
\label{F2}
\end{figure}

\begin{figure}[H]
\centering
\includegraphics[width=9cm]{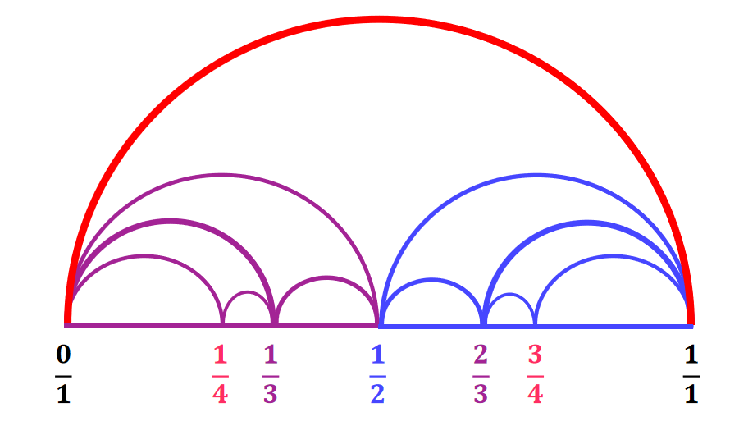}
\caption{The Farey diagram $\mathcal{F}_{4}$ represented with circular arcs
by the fraction values.}
\label{F3}
\end{figure}

To determine the geodesic of the Hawking radiation, we try to find a
geometric that is hidden behind the evolution of states $\mathcal{W}%
\left\vert \psi \right\rangle $ and $\left\vert \psi \right\rangle $. As
illustrated above, we also wish to express the state $\left\vert u\bar{u}%
\right\rangle $ in terms of the number of states acting on the state $%
\left\vert \bar{u}u\right\rangle $. Let us start by considering the three
cases $\left( \alpha ,\beta \right) =\left\{ \left( 0,0\right) ;\left(
0,1\right) ;\left( 1,0\right) \right\} $ Eqs.(\ref{s5},\ref{s6}). Then, the
corresponding states generated by%
\begin{equation}
\left\vert u\bar{u}\right\rangle _{\beta =0}=\frac{N_{-}}{\alpha -N_{+}}%
\left\vert \bar{u}u\right\rangle \text{ \ ,\ \ \ \ \ \ }\left\vert u\bar{u}%
\right\rangle _{\alpha =0}=\frac{\beta -N_{-}}{N_{+}}\left\vert u\bar{u}%
\right\rangle .  \label{s7}
\end{equation}%
A new value of $\alpha $ and $\beta $ is in principle determined by the
normalization: $\left\langle u\bar{u}\right. \left\vert u\bar{u}%
\right\rangle =1$, one could take
\begin{equation}
\alpha =\left\{ \left( N_{+}-N_{-}\right) ;N\right\} _{\beta =0}\text{ \ \
,\ \ \ \ }\beta =\left\{ \left( N_{-}-N_{+}\right) ;N\right\} _{\alpha =0}.
\label{s8}
\end{equation}%
Here, we have $N\neq 0$. To check the conditions Eqs.(\ref{QQ},\ref{s8}), we
distinguish between two cases: \newline
(i) For the two cases: $\alpha =N_{+}-N_{-}$ or $\beta =N_{-}-N_{+}$, the
number of states are $N>1$. (ii) For $\alpha =N$ or $\beta =N$, we have
obtained only $\alpha =\beta =N=1$, i.e. only one state in the system. When $%
N=N_{+}-N_{-}$, one could take one degenerate solution corresponding to $%
N_{-}=0$, which agrees with the violation of CP symmetry between matter and
antimatter \cite{119k}. Therefore, the black hole behaves under the
normalizable states as Schwarzschild or extremal black hole. Obviously, this
distinction between the cases of $\alpha $ and $\beta $, can simplify the
equation Eq.(\ref{s7}) as%
\begin{eqnarray}
\left\vert u\bar{u}\right\rangle &=&\left\vert \bar{u}u\right\rangle \text{
\ for }N=1,  \label{s9} \\
\left\vert u\bar{u}\right\rangle &=&-\left\vert \bar{u}u\right\rangle \text{
\ for }N>1.  \notag
\end{eqnarray}%
Then, the corresponding states generated by
\begin{equation}
\left\vert u\bar{u}\right\rangle =\left( -1\right) ^{1+\delta
_{N}}\left\vert \bar{u}u\right\rangle ,  \label{s10}
\end{equation}%
where $\delta _{N}\left( N=1\right) =1$ and $\delta _{N}\left( N\succ
1\right) =0$. Both cases Eq.(\ref{s7}) give the same result above. These
states are invariant under the transformation $\delta _{N}\rightarrow
-\delta _{N}$. Therefore, the state $\left\vert u\bar{u}\right\rangle $ is
formulated in terms of the state $\left\vert \bar{u}u\right\rangle $, the
main reason is that this relation gives an aspect in of a wormhole between
the two states. It was explicitly confirmed by ER bridges from ER = EPR \cite%
{ER6}. This result is equivalent to the sum of the M\"{o}bius function $\mu
\left( N\right) $ over $N$
\begin{equation}
\delta _{N}=\sum_{k|N}\mu \left( k\right) =\left\{
\begin{array}{c}
1\text{ \ \ if \ }N=1 \\
0\text{ \ \ if \ }N\succ 1%
\end{array}%
\right. \text{ ,}  \label{s11}
\end{equation}%
where%
\begin{equation}
\mu \left( N\right) =\sum_{\substack{ 1\leq k\leq N  \\ \gcd \left(
k,N\right) =1}}e^{2\pi i\frac{k}{N}}.  \label{s12}
\end{equation}%
This algebraic proof shows that the connection between the two horizons $%
\mathcal{H}_{L}$\ and $\mathcal{H}_{R}$ is realised the M\"{o}bius function.
Thus allowing to the Mertens function:%
\begin{equation}
\mathcal{M}_{N}=\sum_{k=1}^{N}\mu \left( k\right) =-1+\sum_{n\in \mathcal{F}%
_{N}}e^{2\pi in},  \label{s13}
\end{equation}%
where $\mathcal{F}_{N}$ denotes the Farey sequence of order $N$ \cite{FS}.
The quantum number $\alpha $ Eq.(\ref{QQ}), consists of a contribution from
the Farey sequence. The number $\alpha $ can be encoded into $\mathcal{F}%
_{2}=\left\{ \frac{0}{1},\frac{1}{2},\frac{1}{1}\right\} $. Let us consider
the fraction $\frac{a}{b}\in \mathcal{F}_{N}$, there is a Ford circle $%
\mathcal{C}\left[ \frac{a}{b}\right] $ with the radius $1/(2b^{2})$ and
centre at $\left( \frac{a}{b},\frac{1}{2b^{2}}\right) $. The Farey diagram
describes a geometry near the AdS horizon. The Farey diagram $\mathcal{F}%
_{N} $ of order $N\longrightarrow \infty $ is given by%
\begin{equation}
\mathcal{F}_{\infty }=\left\{ \frac{a}{b}:\left( a,b\right) \in
\mathbb{Z}
,\text{ }\gcd (a,b)=1\right\} \cup \left\{ \frac{1}{0}\right\} .  \label{s14}
\end{equation}%
The graph $\mathcal{F}_{N}$ is associate to a closed surface of genus. This
represents that the edges indicate geometric intersection number equal to $N$
\cite{P2}. In fact, we hcan also study the possible symmetries of the Farey
diagram using $2\times 2$ matrices correspond to linear transformations of
the plane from linear algebra, see Fig.\ref{F2}.
Each such vertices of $\frac{a}{b}\in \mathcal{F}_{N}$ has two primitive
elements $\left( a,b\right) $ and $\left( -a,-b\right) $. This makes it
possible that $a/b$ and $\left( -a\right) /\left( -b\right) $ are two
different descriptions of the same state. In Fig.\ref{F3} we give an example
of Farey diagram.
\begin{figure}[H]
\centering\includegraphics[width=7cm]{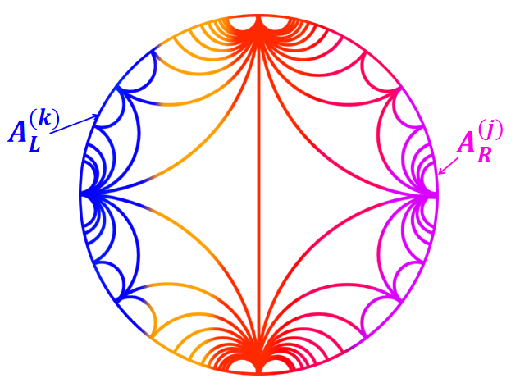}
\caption{The bulk minimal surfaces of the subregions of $\mathcal{A}_{L}$
and $\mathcal{A}_{R}$. In the surface, we see the presence of semi-circles,
which means, in the hrizon there is the presence of edge states. This
corresponds to the quantum hall effect.}
\label{F4}
\end{figure}
The Farey sequences of order $n$ in $CFT=\mathcal{A}_{L}\cup \mathcal{A}_{R}$
circle encode an information contained in a set of coefficients $\frac{a}{b}$%
. This result can be extended to include subregions Fig.\ref{F4}: $\mathcal{A%
}_{L\text{ or }R}^{(n)}\subset \mathcal{A}_{L\text{ or }R}^{(n-1)}\subset
\cdots \subset \mathcal{A}_{L\text{ or }R}^{(1)}\subset ou\mathcal{A}_{L%
\text{ or }R}$. Following this procedure we obtained the $n$-point
correlation function corresponding to the fraction points of the Farey
diagram. Later, we will see a new description of $n$-point correlation
function, like $q$ cell is characterized by $q$-bit code. Each curve in Fig.%
\ref{F4}, represents a geodesic connection entanglement wedge between two
horizon points. This geodesic represents the quantum states of black hole
interior and only the description of the black hole entropy. On the other
hand, there remains the geometric description of the exterior region of the
black hole and the Hawking radiation. If we complete the circles in the
border of the Farey diagram, we will construct the geodesics of the exterior
region
\begin{figure}[H]
\centering\includegraphics[width=11cm]{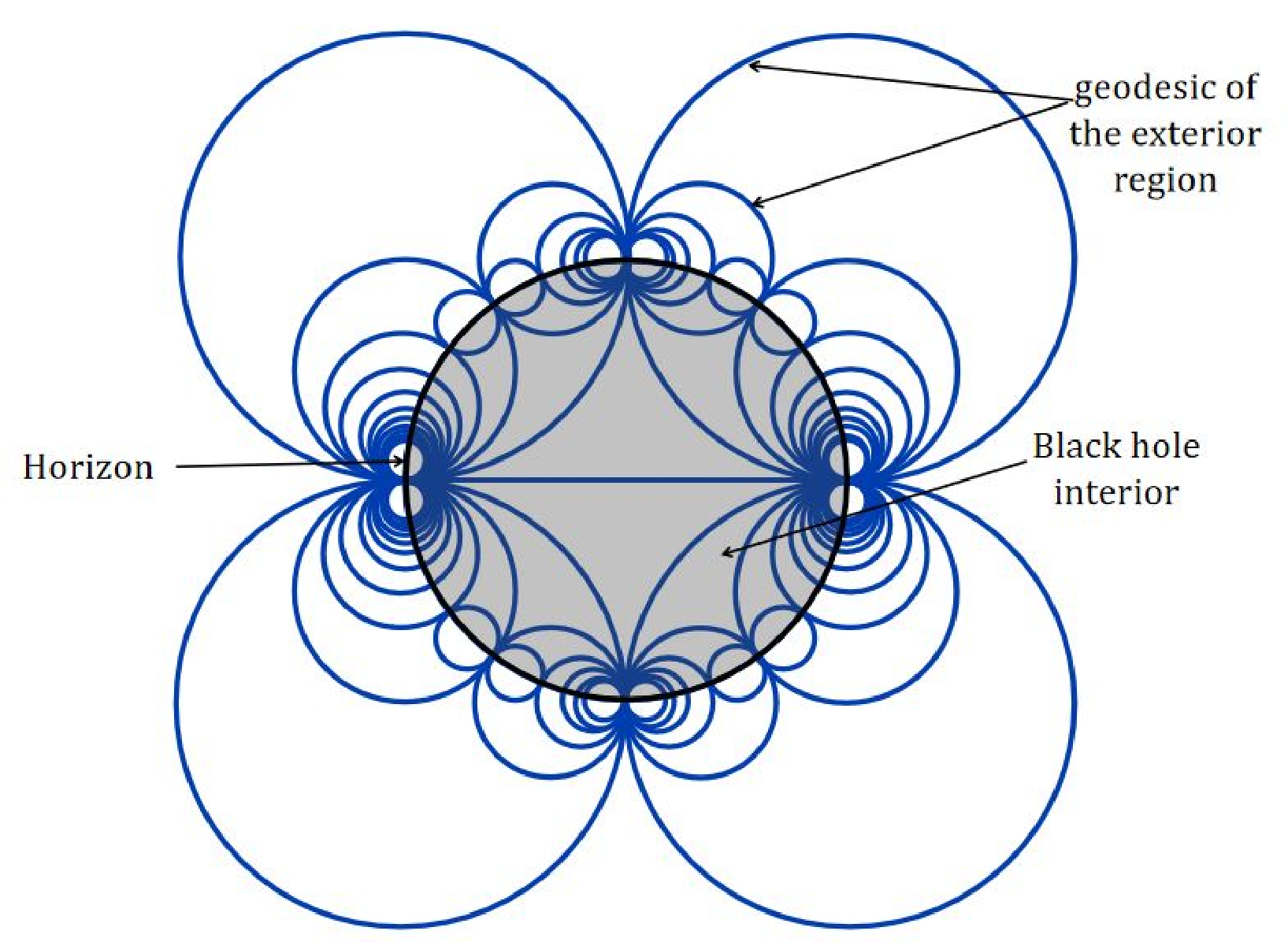}
\caption{To obtain Farey diagram, we employ the Ford circle \protect\cite{WW}%
. The diagram also makes it very clear that there the event horizons in Ford
space. The Ford circle are equivalent to the cylcotron orbit of the edge
states in the near horizon region.}
\label{C}
\end{figure}
The geodesic of Farey diagram in the exterior region is in a good agreement
with the bath coupled to AdS black hole \cite{SQ}, which absorbs the
radiation from the black hole. From Fig.\ref{C}, we identify the curves in
the exterior region as the geodesic of the trajectory of the Hawking
radiation. The permanent creation of radiation Eq.(\ref{F}), is due to the
rotation of radiation in the same circles over time. According to this
geodesic, the radiation does not fall into the black hole due to gravity but
because of the special geodesic of the Farey diagram. The origin of the
radiation geodesics, is the projection of evolution of TFD states in AdS,
since AdS is the dual of CFT. So, the CFT states correspond to the AdS
geodesics. The interaction between the horizon and the Hawking radiation
create coding cells, which can be identified by $\mathcal{A}_{L\text{ or }%
R}^{(n)}$. To complete this geometrical description, one must treat this
geodesic according to the unitary describtion \cite{PP3}.

\section{Unitary prime describtion of AdS geometry}

The Farey diagram of theAdS geometry is a construction of multi-boundary
connection between microstates. The Multi-boundary can be obtained by
evaluating the Euclidean CFT path integral on a cut Riemann surface \cite%
{PP5}. The connection between the circles in the Farey diagram, represent
the encoding of information. The number of connected components of $\mathcal{%
F}_{N}$ is constructed as follows%
\begin{equation}
N_{con}\left( \mathcal{F}_{N}\right) =\left\{
\begin{array}{c}
p^{q-1}\left( p+1\right) \text{ \ \ if \ }N=p^{q} \\
\infty \text{ \ \ \ \ \ \ \ \ \ \ \ \ \ \ }otherwise.%
\end{array}%
\right. ,  \label{s15}
\end{equation}%
where $p$ is a prime and $q$ is a positive integer. When $N$ is not a prime
power, the number of connected components is infinite. We are interested by
the solution where $N=p^{q}$. This indicates that the convergence to the
large $N$ limit is fast. The exponential degeneracy of the field theory
vacuum states, indicated by the number
\begin{equation}
N_{vac}=e^{q\log p},  \label{s16}
\end{equation}%
which corresponds to the number of vacuum states in \cite{PP3} if the
entropy of a vacuum is $S_{vac}=ql_{P}^{2}\log p$, where $l_{P}=\sqrt{\hbar
G/c^{3}}$ is the Planck length. Therefore, the logarithm of the prime number
of independent quantum states describes the boundary structure of the black
hole. This number is also equivalent to the number of black hole states
consistent with the boundary structure described above. We consider a
lattice system in the horizon, each $q$ cell is describe by a Farey fraction
$\frac{a}{b}$. This implies that $S_{vac}$ is obtained only after we include
a prime number $p$. Also $N_{vac}$ represents the number of possible
independent ways to describe CFT on a fixed classical spacetime background
(semi-classical physics). Let us elaborate the entropy of the near horizon
region in a similar way to the Bekenstein-Hawking formula. The black hole
first law of thermodynamics reads $S_{BH}=\int dM/T$. The validity of the
generalized second law of thermodynamics suggests that it is given by the
exponential of the Bekenstein-Hawking entropy $S_{BH}=A/4$ \cite{PP3}, where
$A$ is the area of the horizon. More precisely, the black hole horizon is
encoded on each $q$ cell by the prime number of states
\begin{equation}
p=\exp (A/4ql_{P}^{2}).
\end{equation}%
According to \cite{PP3}, the\ term $\exp (A/4ql_{P}^{2})$ is the the total
number of black hole vacuum states $\left\vert \psi _{q}\right\rangle $ for
a fixed mass, where $q=1,2,...,p$. The entropy $S_{BH}$ will be the encoding
of $p$ for $p$ cell and we write that
\begin{equation}
S_{BH}=pl_{P}^{2}\log p.  \label{P}
\end{equation}%
Therefore, the prime numbers $p$ are always the total number of states of a
black hole. In this view, the state $\left\vert u\bar{u}\right\rangle $
gives the unit cell with area $l_{P}^{2}$. Sometimes the existence of the
connected components of $\mathcal{F}_{N}$ Eq.(\ref{s15}) is presented as a
consequence of the area law. This shows that $\mathcal{F}_{N}$ diagram is a
new geometrization of black hole. In the Hawking radiation process, the
exponent of the Bekenstein-Hawking entropy represented the density of state,
which can be written as an $\left( p\right) ^{pl_{P}^{2}}$. Eq.(\ref{P})
shows that the area of the black hole increases over time in a discrete way
according to the distribution of prime numbers. In unitary describtion of
BH, there is an increasing in the Planck surfaces in the horizon according
to the integer $p$. We notice that from Eq.(\ref{P}), The information of
depends not only on the geometry but also the arithmitic of coding of the
unit cells.

\section{Fermi--Dirac of prime number state}

\begin{figure}[H]
\centering\includegraphics[width=11cm]{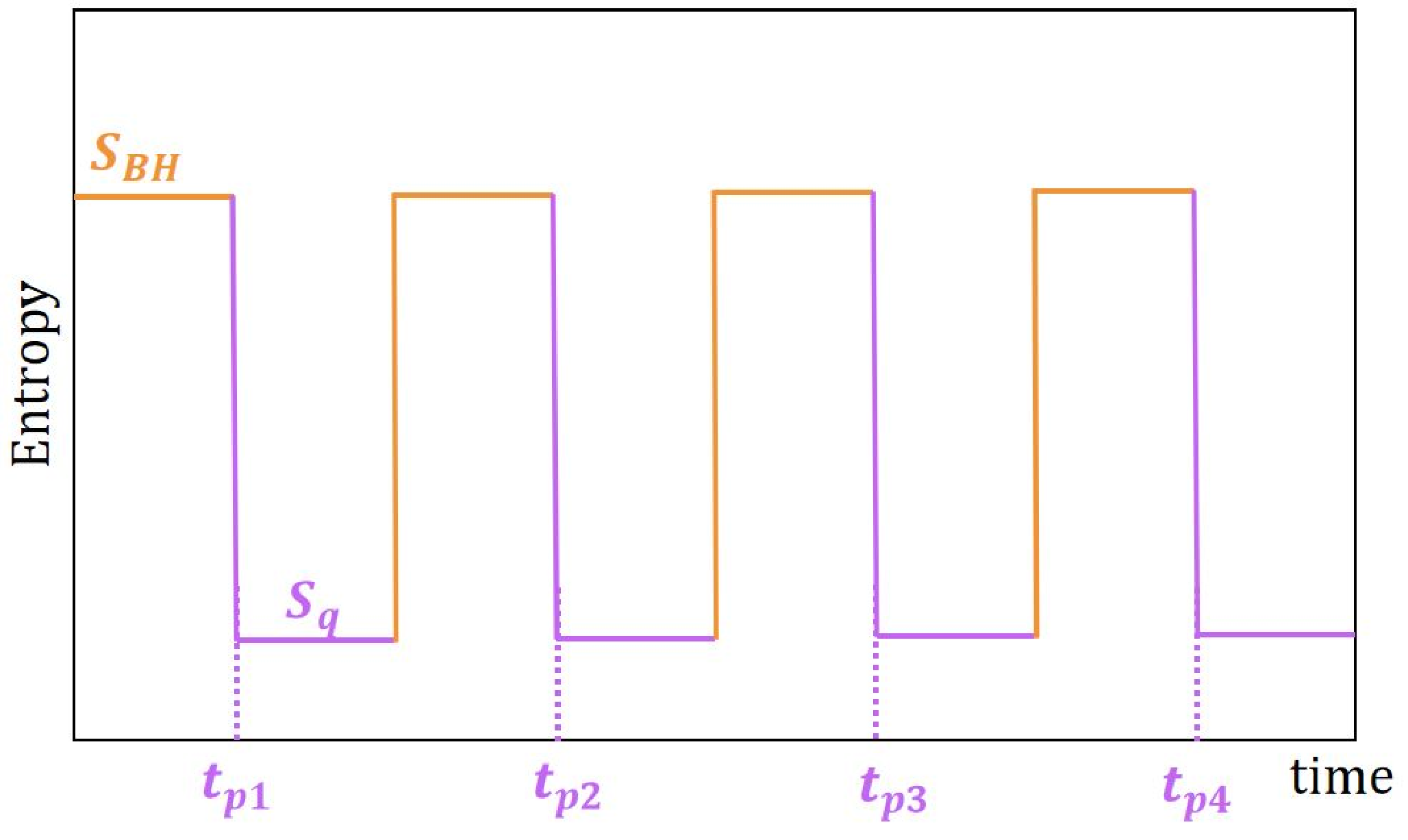}
\caption{This figure shows the periodicity of emission and return of Hawking
radiation in the case of an isolated black hole, which is equivalent to
generalized Page curve.}
\label{SQ}
\end{figure}

Next we build a unitary processes, describing the interior and near horizon
the black hole vacuum states. The number of excited states that can be built
is%
\begin{equation}
N_{tot}=N_{con}=\left( 1+\frac{1}{p}\right) \times N_{vac},  \label{s17}
\end{equation}%
when $p=\infty $ we get $N_{tot}=N_{vac}$. The prime number $p$ represents
the encoding of the information on the horizon. As a result, we find the
Fermi--Dirac particle-energy distribution%
\begin{equation}
N_{vac}=\frac{N_{tot}}{e^{-\frac{S_{BH}}{ql_{P}^{2}}}+1}.  \label{s18}
\end{equation}%
If $q=0$, we obtain $N_{vac}=N_{tot}$. We notice that $N_{vac}$ is athe two
point functions of the $q-$modes \cite{HH5}. We can immediately see that the
simplest horizon corresponds to a thermal ensemble of Fermi--Dirac
configuration, which justifies our focus on the number of connected
components of $\mathcal{F}_{N}$. Therefore, the\ prime power $N$ obtained by
the Farey sequence, generates the Fermi-Dirac statistics. This implies that
we can interpret $S_{q}=ql_{P}^{2}$ as an entropy of $q$ cell of the horizon
with the quantum microstates $\left\vert \psi _{q}\right\rangle $, where $%
q=1,2,...,p$ and $p$ is the total number of black hole states.\newline
The horizon and near horizon region\ are described by the microstate $%
\left\vert \psi _{q}\right\rangle $. The analysis described above indicates
that the vacuum state is defined by $a^{\left( 1\right) }\left\vert \psi
_{1}\right\rangle =0$ and $a^{\left( q\right) }\left\vert \psi
_{q}\right\rangle =\left\vert \psi _{q-1}\right\rangle $,\ where $a^{\left(
1\right) }$ is annihilates all the vacuum states. A state in which describes
the black hole can be constructed by acting $a^{\left( q-1\right) \dagger
}\left\vert \psi _{q-1}\right\rangle =\left\vert \psi _{q}\right\rangle $,
where $a^{\left( q\right) \dagger }/a^{\left( q\right) }$ are the
creation/annihilation operators of order $q$. For $p=2$ we obtain $%
N_{vac}=2^{q}$. The states $\left\vert \psi _{r>n}\right\rangle $
corresponding to the black hole radiation by thermal ensembles of
excitations with the temperatures $1/\beta $. Later, we shall see that the
quantum Hall effect from this entropy. Using Eq.(\ref{s18}), we define the
entropy of the code space $H_{code}\subseteq H_{CFT}$ \cite{SQ} of $q$-cell
in the horizon by
\begin{equation}
S_{q}=l_{P}^{2}\frac{\log N_{vac}}{\log p}.  \label{s19}
\end{equation}%
According to \cite{SQ}, we notice that $S_{p}\sim \log \left\vert
H_{code}\right\vert $. In this case there exists a subspace $%
H_{code}^{\left( q\right) }\subseteq H_{code}$ of $q$ cell ( $q$
microstates). By equivalence, we get $S_{q}\sim \log \left\vert
H_{code}^{\left( q\right) }\right\vert $. The $q$ cell is characterized by $%
q $-bit code \cite{MOS}. So that $\log \left\vert N_{vac}\right\vert ^{\frac{%
l_{P}^{2}}{\log p}}=\log \left\vert H_{code}^{\left( q\right) }\right\vert $%
. We can express the Bekenstein-Hawking entropy and the entropy of a vacuum
of coding entropy in terms of the encoding entropy $S_{q}$ as $%
S_{vac}=S_{q}\log p$ and $S_{BH}=S_{p}\log p$. The presence of $\log p$ in
the entropy $S_{BH}$, implies a discontinuity of the entropy at the level of
the information coding.%
\begin{equation}
S_{BH}/l_{P}^{2}=p\times \log p\text{ \ \ ;\ \ \ \ }S_{q}/l_{P}^{2}=\log
N\times \frac{1}{\log p}  \label{ss19}
\end{equation}%
By comparing the black hole entropy Eq.(\ref{P}) with the coding entropy Eq.(%
\ref{s19}). We notice that when there is an increase in $p$-cell over time,
there is an increasing in $S_{BH}$, and a decrease in $S_{q}$. Therefore,
during the increase of the BH area, the encoding entropy decreases, which
shows that the entropy $S_{q}$ is the von Neumann entropy of the Hawking
radiation \cite{PA1}. Since we are studying the the black hole in AdS
background, where there is the presence of time. If we assume that $S_{q}$
is the von Neumann entropy of the Hawking radiation. The evolution of $p$%
-cell is proportional to time $t$, in this case we propose that $\mathcal{P}%
=\log p\sim t$.
\begin{figure}[H]
\centering\includegraphics[width=11cm]{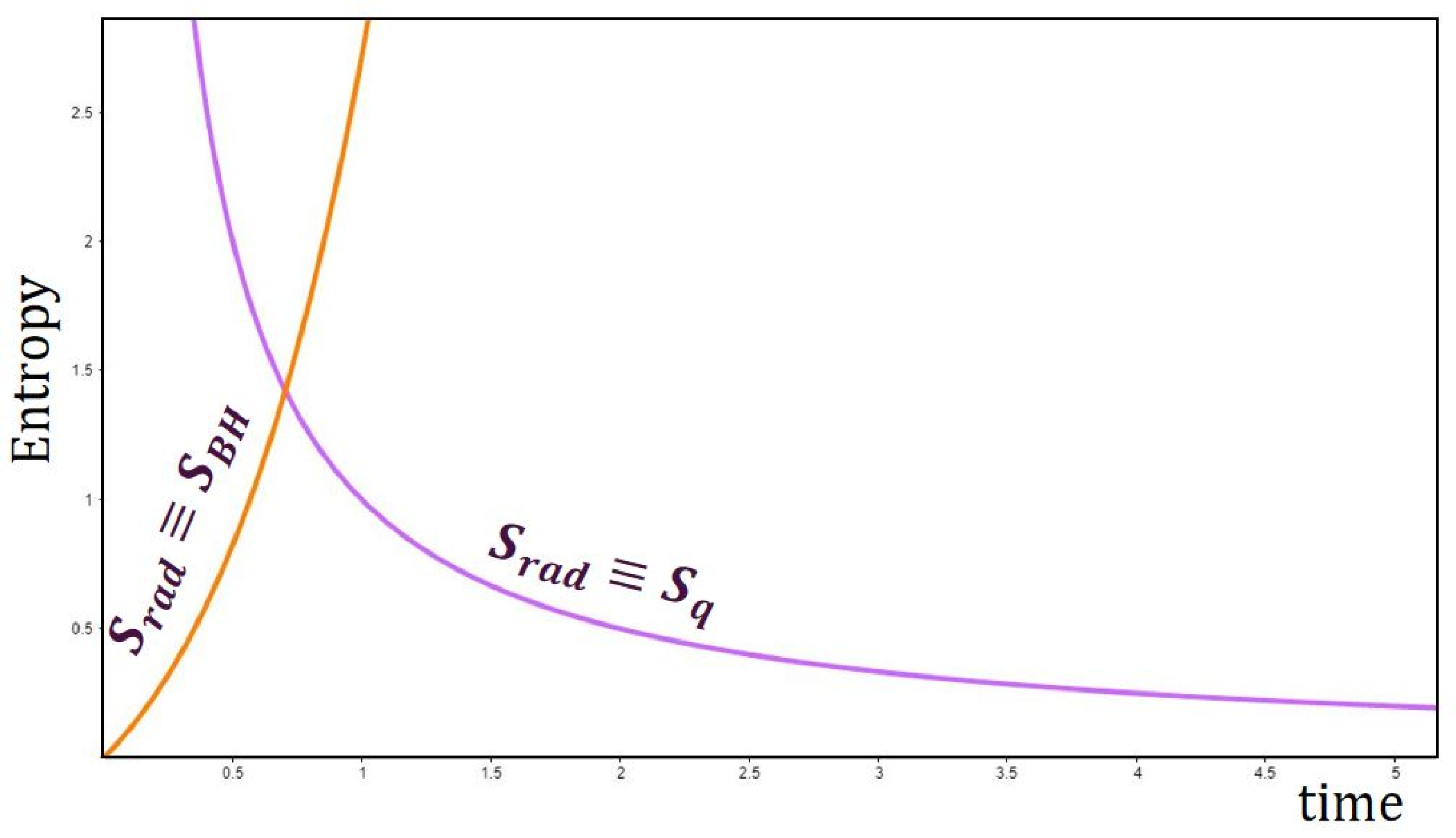}
\caption{This figure shows the behaviors of $S_{BH}/l_{P}^{2}=\mathcal{P}e^{%
\mathcal{P}}$\ dashed line and $S_{q}/l_{P}^{2}=\frac{\log N_{vac}}{\mathcal{%
P}}$ dashed line in terms of $\mathcal{P}$. \ We suppose that $\mathcal{P}$
is continuous.}
\label{S}
\end{figure}
According to the Fig.\ref{S}, and the two equations (\ref{ss19}), the
quantum state follows $\log p$ inside black hole, and they come out of the
black hole in the form of radiation. Outside the black hole, they follow $%
1/\log p$, which changes the direction of these radiation, then they fall on
the horizon. On the other hand, there is a problem, it is that the change of
entropy $S_{BH}$ (interior) to $S_{q}$ (exterior), must occur in the Page
time. But we see that there is a return of the radiations in the black hole
to another emission, indeed there are always emissions and returns of these
radiations Fig.\ref{SQ}. To solve this problem, there are very important
indications in the entropies Eq.(\ref{ss19}), these entropies are discrete
according to the prime number $p$. This shows that there is emission of
radiation in an interval of time, until the first page time, then there is a
return of these radiation to the black hole according to the geodesics of
the Farey diagram. When all the radiations return, it proceeds to the second
emession, and the same senario is always repeated. To see things more
clearly we use some numerical values:\newline
(1) the first emission of radiation according to $\log 2$, until the page
time $t_{p2}$, then it returns to the black hole according to $1/\log 2$.%
\newline
(2) the second emission of radiation according to $\log 3$, until the page
time $t_{p3}$, then it returns to the black hole according to $1/\log 3$.%
\newline
(3) the third emission of radiation according to $\log 5$, until the page
time $t_{p5}$, then it returns to the black hole according to $1/\log 5$.

\begin{equation*}
\vdots
\end{equation*}%
(p) the $p$-th emission of radiation according to $\log p$, until the prime
page time $t_{pp}$, then it returns to the black hole according to $1/\log p$%
.
The emission of the Hawking radiations occurs in a discrete time $t\sim \log
p$ and are returned is done in the reverse of this discrete time. When
Hawking radiation falls on the horizon, he will have the production of a
resistance which resists radiation from entering the black hole. For that we
will study in the next section the type of this resistance and its influence
in this radiations.

\section{Quantum Hall effect in the horizon}

To explain the quantification of the entropy values and the other physical
properties in this model, we propose to study the the Hall resistance in the
black hole. Because we know that in the quantum Hall models, the values of
the Hall resistance are quantified. Also because we have already seen in Fig.%
\ref{F4} that there is the presence of the semi-circles in the edge of Farey
diagram, which indicates the edge states. Using the black hole first law of
thermodynamics for semiclassical black holes reads $S_{BH}=\int dM/T=A/4$.
One could also choose $S_{p}\equiv \int dM/T_{p}$. In the simple case, we
define the resistance $R_{H}$ which depends on the temperature of $p$ cell as%
\begin{equation}
R_{H}\left( q\right) =\frac{4\pi \varepsilon _{0}k_{B}T_{p}\log p}{c^{3}M}%
\equiv \frac{T_{p}\log p}{M},  \label{ss20}
\end{equation}%
or equivalently%
\begin{equation}
R_{H}\left( p\right) =\frac{1}{l_{P}^{2}p}.  \label{s20}
\end{equation}%
We notice that the expression Eq.(\ref{s20}) is exactly like with the Hall
resistance $R_{H}$ in the quantum Hall effect, because $p$ is integer: $%
R_{H}=h/e^{2}\nu =1/G_{H}$, where $h$ is Planck's constante, $e$ is the
elementary charge and $G_{H}$ is the quantized Hall conductivity. The number
$\nu $ can take on either integer $(\nu =1,2,3,...)$ or fractional $\nu
=(1/3,2/5,3/7,...)$. It is a simple consequence of the motion of charged
particles in the near horizon region. In this case the entropy Eq.(\ref{s19}%
) represent the quantized Hall conductivity \cite{HH3}, i.e. $l_{P}$ is the
elementary area of the elementary charge $e$. The most important remark is
when the number of $p$ cells is infinite, the resistance $R_{H}$ is zero,
then there is no more evaporation of the black hole ($T_{\infty }=0$).
\newline
From Eqs.(\ref{s19},\ref{ss20},\ref{s20}) we get
\begin{equation}
S_{p}\equiv G_{H}\propto \frac{1}{T_{p}}.  \label{s21}
\end{equation}%
This last relation is similar to \textbf{Wiedemann-Franz law} \cite{HH4},
i.e. the electrical conductivity of metals at normal temperatures is
inversely proportional to the temperature. Also we have $R_{H}\left(
p\right) =1/S_{BH}\sim 1/A$, which checks the law of the \textbf{Pouillet's
law }\cite{HH5}. Recall that the quantum Hall is observed in two-dimensional
electron systems. This result corresponds exactly with area law, which
connects the entropy of black hole with the surface of horizon. We notice
that the relation between area law and the entropy $S_{q}$, is described by
the relation Eq.(\ref{s18}). The electrical resistance in the Pouillet's law
can be expressed as
\begin{equation}
R_{H}=\frac{r}{A\sigma },  \label{s22}
\end{equation}%
where $A$ is the cross-sectional area, $r$ is a distance between the horizon
and the charged near horizon region, $\sigma $ is the conductivity of the
matter (material) in the near horizon region. From Eqs.(\ref{s20},\ref{s22}%
), we obtain \newline

\begin{equation}
A\sigma =prl_{P}^{2}.
\end{equation}%
If the total number of cells equal to 1; i.e. $n=A/4l_{P}^{2}=1$, we find $%
\sigma _{1}=\frac{r}{4}$. If we assume that $r$ is the thickness of the
region between the black hole and the photon sphere, we can express in this
case $\sigma $ as $\sigma _{1}=\frac{1}{4}\left( r_{ps}-r_{H}\right) $,
where $r_{ps}$ is the radius of the photon sphere. In this region, we have
electriclly charged particle intercting with electromagnetic field. Then, we
propose a representation of the near horizon geometry in terms of magnetic
field $B$. We have a creation of a magnetic field on each cell with the
magnetic length $l_{B}$, which will be proportional to Planck length $l_{P}$%
. We can interpret this result by creating of an electric current passing
through the horizon in a magnetic field. \newline
From Fig.\ref{C} and Fig.\ref{F3}, we can use the quotients of the Poincar%
\'{e} disk \cite{PP4} to construction of the wormhole spatial geometries.
From Farey graph Fig.\ref{C}, we notice that the circles are only the
displacement of charges on the black hole surface, and the interior of the
black hole behaves as a topological insulator. In a classical point of view,
the charged particles move in circular motion (the cyclotron orbit) in the
horizon with a uniform perpendicular magnetic field. This rotational motion
predicts the existence of edge states in the the black hole horizon. This
concept makes it possible to see the lines are equipotential lines.

\section{ Discussion}

(i) \textit{Summary.}\newline
\emph{This paper,} based on the geometrization of the near-horizon region.
We noticed that there is a new geometry hiding in the TFD states. In this
framework we have defined and studied new states expressed as a function of
TFD state and for an infinite temperature. We have divided the horizon into
two sub-regions; the left horizon $\mathcal{H}_{L}$\ and the right horizon\ $%
\mathcal{H}_{R}$. Our results clearly demonstrate that the connection
between the two horizons $\mathcal{H}_{L}$\ and $\mathcal{H}_{R}$ is
realised by the M\"{o}bius function. The evolution of the TFD state in the
horizon allows to distinguish between two cases of the number of particles $%
\left\{ N=1,N\succ 1\right\} $. We have geometrized the near horizon region
by the Farey diagram $\mathcal{F}_{N}$. The use of the number of connected
components of $\mathcal{F}_{N}$, leads to describe unitary cells of the
horizon. This description is done by the Fermi--Dirac distribution. From
this distribution, we have demonstrate the existence of cells that contain
quantum information. The number of connected components in Farey diagram
encode the quantum information in cells. The $q$ cell is characterized by $q$%
-bit code and the prime number of states $p$. Each $q$ cell is described by
the entropy $S_{q}=ql_{P}^{2}$ and the quantum microstates $\left\vert \psi
_{q}\right\rangle $. If $q$ equals the total number $p$ (prime number) of
cells in the horizon, the Bekenstein-Hawking entropy will be equal to $%
pl_{P}^{2}\log p$. Following this procedure, we have write the
Bekenstein-Hawking entropy in logarithmic term. We have found the expression
of the horizon resistance in terms of $1/p$ and which is exactly similar to
the Hall resistance $R_{H}$ in the quantum Hall effect. The formulation of
this resistance leads to a new description of the black hole evaporation.
\emph{As a brief summary}, we has used the relationship between geometry and
number theory to describe the encoding of quantum information on the horizon
and black hole evaporation, which shows us that the black hole behaves like
a topological insulator. The geometry of \ CFT states, is represented by the
Farey diagram, behind this diagram it has a physics description, this is a
unitary describtion of quantum Hall effect in the Fermi--Dirac statistics.%
\newline

(ii) \textit{Radiation journey.}\newline
During the increase of the black hole entropy according to the quantum
information of the CFT states, there is a production of the traces of
information in AdS background. This connection intrinsic connection between
geometry and informationis already known in holographic principle \cite%
{HP1,HP2}. The traces of the information in AdS geometry are described by
the geodesics and the lines of the Farey diagram. After the first Page time,
there is the beginning of the radiations emission. This radiations follow
the geodesic of AdS- Farey diagram then they fall in the horizon. During the
transfer of radiation to information inside the black hole, the horizon
resists the passage according to quantum Hall effect, this scenario always
repeats. The increase of the black hole entropy implies the increase of the
geodesic circles of the radiation in the exterior region. This geodesic of
the exterior region represents the bath coupled to the AdS black hole. If
the black hole is not isolated, and it still absorbs matter and energy, its
area increases also and the Farey geodesic region also increases. In this
case, the radiation travel through a long path before they return to the
black hole.

\end{document}